\documentclass[aps,prl,preprint,groupedaddress,showpacs]{revtex4}

\usepackage{graphicx}
\usepackage{epsfig}

\usepackage[english]{babel}
\newcommand{\etal}{\emph{et al.}}
\newcommand{\ie}{\emph{i.e.}}

\begin{document}

\preprint{Physical Review Letters, in press}

\title{Exciton Regeneration at Polymeric Semiconductor Heterojunctions}

\author{Arne C.\ Morteani}
\author{Paiboon Sreearunothai}
\author{Laura M.\ Herz}
\altaffiliation[Current address: ]{Clarendon Laboratory, University of Oxford, Parks Road, Oxford OX1 3PU, United Kingdom}
\author{Richard H.\ Friend}
\author{Carlos Silva}
\email[Corresponding author. E-mail: ]{cs271@cam.ac.uk}
\affiliation{Cavendish Laboratory, University of Cambridge, Madingley Road, Cambridge CB3 0HE, United Kingdom}

\date{\today}

\begin{abstract}

Control of the band-edge offsets at heterojunctions between organic semiconductors allows efficient operation of either photovoltaic or light-emitting diodes. We investigate systems where the exciton is marginally stable against charge separation, and show via E-field-dependent time-resolved photoluminescence spectroscopy that excitons that have undergone charge separation at a heterojunction can be efficiently regenerated. This is because the charge transfer produces a geminate electron-hole pair (separation 2.2--3.1\,nm) which may collapse into an exciplex and then endothermically ($E_{A}=100$--200\,meV) back-transfer towards the exciton.

\end{abstract}




\maketitle

\begin{figure}
    \includegraphics[width=7.5cm]{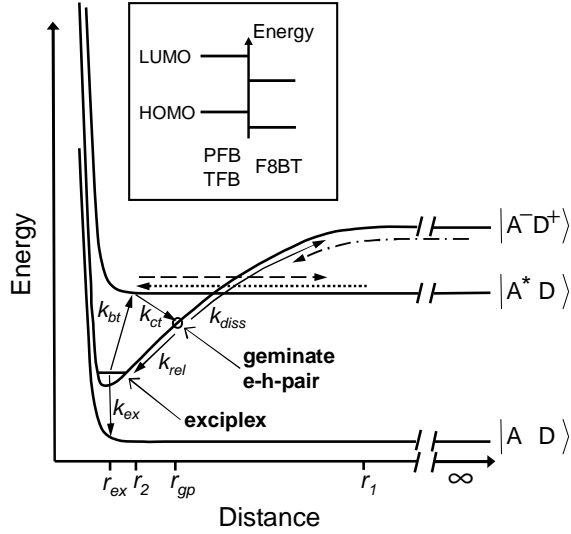}%
    \caption{\label{fig:Model} Potential energy diagram describing the energetics and kinetics at type II polymer heterojunctions. The energetic order of $\mid$A$^{-}$D$^{+}\rangle_{r=\infty}$ and $\mid$A$^{*}$D$\rangle_{r=\infty}$ may be reversed for PFB:F8BT vs. TFB:F8BT. The inset shows the band offsets at a type II heterojunction (see also \cite{Morteani.Dooht.ea:03}).}
\end{figure}

Efficient optoelectronic devices fabricated with semiconductor polymers
often employ heterojunctions between two components in which both the
electron affinity and ionization potential are higher in one
material than in the other (`type II'
heterojunctions, see inset of Fig.~\ref{fig:Model}). This configuration is commonly used in
photovoltaic diodes to achieve charge generation at the hetero-interface~\cite{Halls.Walsh.ea:95,Arias.MacKenzie.ea:01,Brabec.Sariciftci.ea:01}. Typical devices involve a thin film of a blend of hole-accepting and electron-accepting polymers
sandwiched between two electrodes. However, some type II
polymer blends show low photocurrents and high luminescence
quantum yields, leading to very efficient light-emitting
diodes~\cite{Halls.Cornil.ea:99,Cao.Parker.ea:99,Palilis.Lidzey.ea:01,Morteani.Dooht.ea:03}. 

The high luminescence quantum yield is commonly rationalized by
the proposition that excitons can be stable at the heterojunction
if their Coulombic binding energy is higher than the band edge
offsets~\cite{Halls.Cornil.ea:99}. In this case, the only process that might occur when an exciton encounters the heterojunction is energy transfer from the
material with the larger band gap to the other component. This
picture classifies type II heterojunctions into those above and
those below a charge-separation threshold, producing high
photocurrents or luminescence quantum yields, respectively.
This simple classification is incomplete because even systems
that show high luminescence efficiencies often also show
significant charge generation (see below). By considering the dependence of photoluminescence spectra and dynamics on applied electric field, we
develop here an alternative, unified description of the excitation dynamics at
the polymer heterojunction. We show that in \emph{all} blends the exciton first dissociates at the heterojunction and forms an interfacial geminate charge pair. However, geminate pair recombination via an intermediate heterojunction state (termed an exciplex) can regenerate the bulk exciton. These circular transitions between the different excited states at the heterojunction are driven by thermal energy, and a fine balance of the kinetics determines the net charge separation and photoluminescence yields. 

\begin{figure}
    \includegraphics{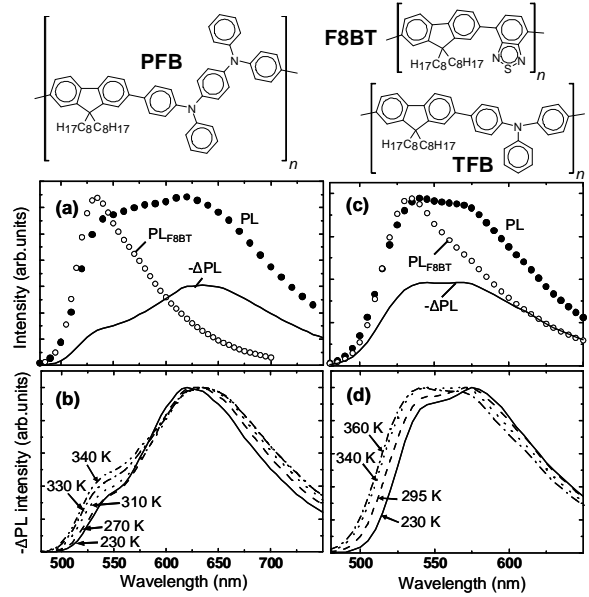}%
    \caption{\label{fig:QuenchingSpec}(a) Photoluminescence intensity (PL, solid circles) and reduction of photoluminescence intensity due to an applied reverse bias of 10\,V ($\Delta$PL, continuous line) for a PFB:F8BT blend device at 340\,K. PL and $-\Delta$PL are plotted in the same scale and reflect their relative intensities. (b) $\Delta$PL spectra (at 10\,V) from the same device as in (a) at different temperatures. (c) PL (solid circles) and $\Delta$PL at a reverse bias of 15\,V (continuous line) for a TFB:F8BT blend device at 340\,K. (d) $\Delta$PL spectra from the same device as in (c) at different temperatures. For comparison the PL spectrum from an F8BT-only device (open circles) is plotted in both part (a) and (c). The structures of PFB, F8BT, and TFB are also shown.}
\end{figure}

Bulk excitons show relatively strong Coulombic binding (of order 0.5\,eV~\cite{Alvarado.Seidler.ea:98,Halls.Cornil.ea:99}), and can be trapped at the heterojunction, acquiring some charge-transfer character. Such excitations are termed exciplexes when seen in isolated donor-acceptor systems and are characterized by featureless, red-shifted emission spectra and long radiative decay times~\cite{Weller:75}. Recently, we have shown that exciplex states form in
blends of F8BT with PFB, and F8BT with TFB (see Fig.~\ref{fig:QuenchingSpec} for molecular structures), and that these exciplex states can undergo endothermic energy transfer to form a bulk F8BT exciton~\cite{Morteani.Dooht.ea:03}. Here we investigate films of PFB:F8BT  and TFB:F8BT spin-coated from common chloroform solution. In general, there is substantial de-mixing of the two polymers through spinodal decomposition during drying, but under the rapid drying conditions achieved here there is more limited de-mixing (of the order of 10\,nm~\cite{Arias.MacKenzie.ea:01}) resulting in a large interfacial area of contact between the two polymers. Note that PFB:F8BT blends can display high charge separation
yields (4\% photocurrent external quantum efficiency) and low EL
efficiencies ($< 0.64$\,lm/W) whereas the TFB:F8BT system displays
low photocurrents (we find 82\% lower short-circuit current than in PFB:F8BT at 457\,nm excitation), but very high electroluminescence efficiencies
(up to 19.4\,lm/W)~\cite{Snaith.Arias.ea:02,Morteani.Dooht.ea:03}. Hence, these blends are good examples for the
contrasting properties of type II polymer heterojunctions as described above.

For all measurements, polymer blends (mass ratio 1:1) were
spin-coated from common chloroform solution onto
oxygen-plasma-treated ITO substrates to form 170\,nm thin films.
Ca electrodes (60\,nm) were then deposited by thermal
evaporation and encapsulated by a 300\,nm Al
layer. All devices were fabricated under N$_{2}$ atmosphere. 
An electric field was applied by reverse-biasing the device to
prevent charge injection (ITO negative with respect to Ca). Quasi-steady-state photoluminescence (PL)
quenching measurements were taken by exciting the sample with a CW
Ar$^{+}$ laser (457\,nm) through the ITO. The resulting
PL was imaged through a monochromator onto a
Si photodiode. A modulated voltage was applied to the device
and changes in PL due to the applied electric field,
$\Delta$PL, were detected using a lock-in amplifier referenced to
the modulation frequency (225\,Hz). The total
PL intensity was measured by mechanical modulation of the laser
excitation. The results reported here are independent of modulation frequency and excitation power. Time-resolved PL measurements were also
performed using time-correlated single photon counting (TCSPC) and
photoluminescence up-conversion (PLUC) spectroscopies with 70\,ps and 300\,fs time-resolution, respectively. Our
TCSPC and PLUC setups are described elsewhere~\cite{Morteani.Dooht.ea:03,Hayes.Samuel.ea:95}. All measurements were taken in
continuous-flow He cryostats (Oxford Instruments OptistatCF)
under inert conditions. Finally, PL efficiency measurements were
performed on simple polymer thin films spin-coated on Spectrosil
substrates using an integrating sphere coupled to an Oriel
InstaSpec IV spectrograph and excitation with the
same Ar$^{+}$ laser as above.

\begin{figure}
    \includegraphics{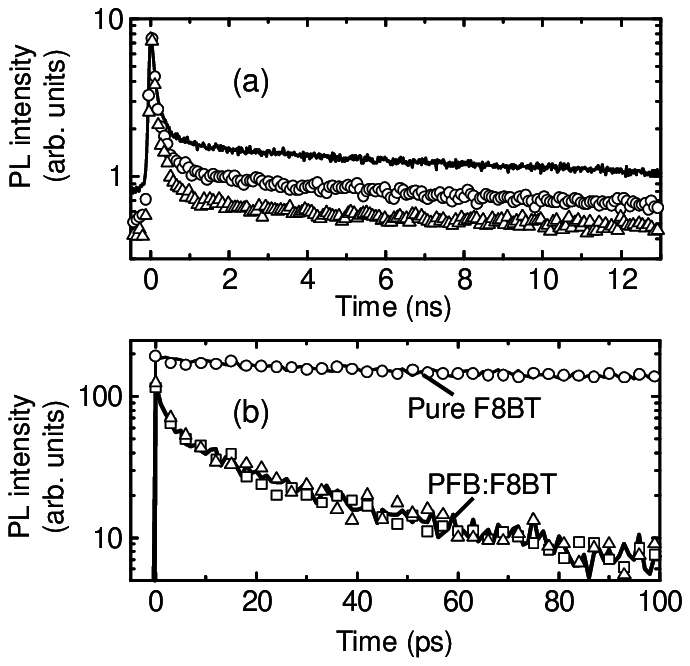}%
    \caption{\label{fig:QuenchingDynamics}(a) Photoluminescence decay measured using TCSPC (excitation: 407\,nm, $<4\,$nJ$/$cm$^{2}$, detection: 640\,nm) from a PFB:F8BT device at room temperature under 0\,V (continuous line), 13\,V (circles) and 30\,V (triangles) applied reverse biases. (b) PLUC measurements (excitation: 405\,nm, $ 42\,$nJ$/$cm$^{2}$, detection: 550\,nm) from a similar device at 0\,V (continuous line), 5\,V (squares) and 12.5\,V (triangles) reverse bias. For comparison, data for a device with pure F8BT at 0\,V (continuous line) and 12\,V (circles) are also plotted.}
\end{figure}

Fig.~\ref{fig:QuenchingSpec}(a) compares the PL spectrum of a
diode made with blended PFB:F8BT with that of pure F8BT.
Red-shifted exciplex emission, in addition to bulk F8BT contribution (\ie~the F8BT-only spectrum), is evident in
the blend film. (Neither PFB nor TFB are excited at 457\,nm~\cite{Snaith.Arias.ea:02}.) Also shown in the same figure
is the $-\Delta$PL spectrum taken by applying 10\,V bias across
the device. The electric field preferentially quenches the exciplex contribution in the red part of
the spectrum ($>50$\% quenching for wavelengths $>650$\,nm). Quenching of the F8BT exciton emission is also observed, but decreases with decreasing temperature, as demonstrated in
Fig.~\ref{fig:QuenchingSpec}(b). Similar phenomena are observed in
the TFB:F8BT diode (Figs.~\ref{fig:QuenchingSpec}(c) and
\ref{fig:QuenchingSpec}(d)), although the relative contribution of
F8BT bulk emission is higher in the same temperature range. In contrast to the blends, pure F8BT does not show PL quenching (integrated $\Delta$PL/PL$ \ll$ 1\%) and only Stark-shifts by $<1$\,nm at these fields.

If the PL quenching arises from field-assisted dissociation of
an emissive state, its luminescence decay rate should be
field-dependent. Fig.~\ref{fig:QuenchingDynamics}(a) shows TCSPC
measurements at 640\,nm in a PFB:F8BT diode with different applied
voltages. All curves consist of an instrument-limited decay, and a slow, roughly mono-exponential decay with $40
\pm 5$\,ns time constant. The two components are assigned to the bulk exciton and the exciplex state, respectively~\cite{Morteani.Dooht.ea:03}. Exciplex generation occurs within $\sim 1$\,ns and its generation efficiency is strongly field-dependent, while its decay constant shows no significant variation with applied field. Therefore, an exciplex precursor must be quenched by the field. To investigate the field
dependence on the bulk exciton decay rate, we have performed
field-dependent PLUC measurements. The results are displayed
in Fig.~\ref{fig:QuenchingDynamics}(b). The exciton decay dynamics
are not field dependent \cite{LanzaniPaper}. Therefore, a dark intermediate state must be dissociated by the field. We postulate that this state is an interfacial geminate polaron pair that follows charge transfer from the bulk exciton~\cite{Yokoyama.Endo.ea:81,Wu.Conwell:98,Ohta:02,Arkhipov.Heremans.ea:03}.

\begin{figure}
    \includegraphics{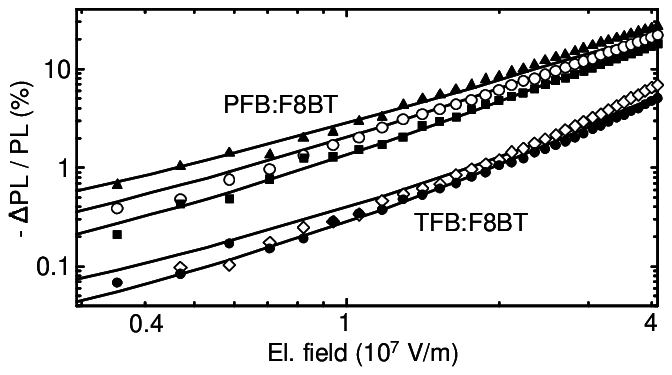}%
    \caption{\label{fig:Onsager}Relative electric field quenching of the PFB:F8BT and TFB:F8BT exciplex photoluminescence intensities (measured at 700\,nm and 580\,nm, respectively), in the same devices as in Fig.~\ref{fig:QuenchingSpec}, versus electric field at 230\,K (solid squares), 250\,K (open and solid circles), 290\,K (solid triangles) and 295\,K (open diamonds). The solid lines through the data are Onsager simulations (parameters for PFB:F8BT: $\epsilon=3.5$, $r_{gp}=3.0$\,nm at $T=230$\,K and 3.1\,nm at 250\,K and 290\,K; for TFB:F8BT: $\epsilon=3.5$, $r_{gp}=2.3$\,nm at $T=250$\,K and 2.2\,nm at 295\,K).}
\end{figure}

To estimate the electron-hole separation within this geminate pair, $r_{gp}$,
we have investigated the field-dependent changes in PL intensity
(Fig.~\ref{fig:Onsager}). Neglecting the effects of energetic disorder~\cite{Nikitenko.Hertel.ea:01} and of a possible interfacial dipole layer~\cite{Arkhipov.Heremans.ea:03}, geminate-pair dissociation in
electric fields is most easily described within the Onsager model~\cite{Pope.Swenberg:99}, which yields the
dissociation probability $f_{\epsilon}\left(r_{gp}, T,
F\right) = f(F)$ of bulk geminate pairs in a medium with dielectric
constant $\epsilon$, under an applied field $F$ and at temperature
$T$~\cite{OnsagerEquations}. 
Since the only material parameter is the dielectric constant, which we approximate to be 3.5 for all polymers, Onsager theory should be applicable also to geminate pairs at the interface.
The field-dependent relative reduction of the geminate pair population $n_{gp}$ is given by $-\frac{\Delta n_{gp}}{n_{gp}}=\frac{f(F) -
f(0)}{1 - f(0)}$. Fig.~\ref{fig:Onsager} plots
$-\frac{\Delta PL}{PL}$ versus electric field\,\cite{InternalField} at various temperatures for
PFB:F8BT and TFB:F8BT devices (measured in the red part of the spectrum where exciton emission is insignificant). Plotted in
the same graph are simulations of $-\frac{\Delta n_{gp}}{n_{gp}}$ using a
$\delta$-function distribution for $r_{gp}$. The simple model fits the data satisfactorily, which supports the assumption
of a geminate pair intermediate prior to exciplex formation and
yields $r_{gp}\approx 3.1$\,nm (PFB:F8BT) and $r_{gp} \approx 2.2$\,nm (TFB:F8BT). The large separation is probably caused by polaron-pair thermalization
following the initial charge-transfer step~\cite{Yokoyama.Endo.ea:81, Pope.Swenberg:99}.

We now return to the $\Delta$PL spectra in
Fig.~\ref{fig:QuenchingSpec}, which contain bulk F8BT components
that are not due to electric-field promoted dissociation of those states as was shown above. The zero-field steady-state
photoluminescence is due to three different excited-state
populations: (\emph{i}) ``primary'' excitons, generated in the
bulk by the laser excitation; (\emph{ii}) exciplexes, generated
via energy transfer from bulk excitons; and (\emph{iii})
``secondary'' excitons, generated via endothermic back-transfer
from the exciplexes \cite{Morteani.Dooht.ea:03}. Since the exciplex density is reduced by application of an electric field, there is less secondary exciton generation, and hence the observed $\Delta$PL contains an excitonic contribution. Further evidence for this hypothesis is provided by the
temperature dependence of the $\Delta$PL spectra shown in
Figs.~\ref{fig:QuenchingSpec}(b) and \ref{fig:QuenchingSpec}(d)~\cite{TemperatureArgument}.
The ratio of secondary excitons to exciplexes is found to follow
an Arrhenius function with activation energy $200 \pm 50$\,meV
(PFB:F8BT) and $100 \pm 30$\,meV (TFB:F8BT). These activation
energies are consistent with those values extracted with our
previous TCSPC measurements~\cite{Morteani.Dooht.ea:03}.

Fig.~\ref{fig:Model} summarizes the above findings. The
potential energy curves represent the ground state
($\mid$AD$\rangle$), the exciton residing on F8BT
($\mid$A$^{*}$D$\rangle$), and the electron and the hole residing
in the respective component across the heterojunction
($\mid$A$^{-}$D$^{+}\rangle$), where A and D symbolize the
acceptor (F8BT) and the donor (PFB or TFB), respectively. The
abscissa represents the intermolecular distance, i.e.
\emph{either} the distance of the exciton from the interface \emph{or} the
separation of the geminate polarons. The exciplex state is then
located in the minimum of the $\mid$A$^{-}$D$^{+}\rangle$
potential. When the system is photoexcited, an exciton is
generated at a certain distance $r_{1}$ from the heterojunction.
It then diffuses to a separation $r_{2}$ (dotted arrow), where it dissociates and an interfacial geminate
electron-hole pair is formed with rate constant $k_{ct}$. This
geminate pair can either dissociate ($k_{diss}$) or relax into the
luminescent exciplex state ($k_{rel}$). The ratio $k_{diss} /
k_{rel}$ is strongly field dependent and determines the degree of
luminescence quenching. The exciplex state can then either decay
($k_{ex}$), or back-transfer to a bulk exciton in F8BT ($k_{bt}$),
but is itself too strongly bound to dissociate under the field. We note that the transition from geminate pairs to excitons via $k_{rel}\rightarrow k_{bt}$ represents a novel mechanism for geminate pair recombination at polymeric heterojunctions. The secondary excitons produced might enter
the cycle again, or diffuse away from the heterojunction (dashed
arrow) and decay. The model is also applicable to electrical
excitation, where the excited state is produced via charge
injection (dash-dotted arrow). The regeneration of the exciton via the thermally-driven circular process $k_{ct}\rightarrow k_{rel}\rightarrow k_{bt}$
means that even though charge transfer occurs, the excitation
energy might eventually still be emitted in the form of bulk
exciton luminescence.

An estimate of the contribution of the regeneration process to the
PL of the blend can be derived by normalizing
the $\Delta$PL spectrum to the PL spectrum at higher wavelengths
where the emission is solely due to exciplexes. We assume that this re-normalized $\Delta$PL then represents the contribution of exciplex and secondary exciton emission to the total PL. We infer thereby that at room temperature in the PFB:F8BT blend $\sim 20$\%
of the visible emission comes from primary excitons. In TFB:F8BT we find a primary exciton contribution of $<2$\%, which implies that $> 98$\% of the excitons undergo  charge transfer at a heterojunction. Despite that, the relative PL quenching with respect to pure F8BT is only $<57$\% (PL yield of F8BT 80\%, of TFB:F8BT 35\%) indicating the great importance of the exciton regeneration mechanism. Secondary exciton and exciplex emission maintain a high PL yield in spite of most excitons encountering a heterojunction. On the other hand, the PFB:F8BT PL yield is only 4\%, consistent with large geminate pair dissociation and low back-transfer efficiency, i.e. with low ``regeneration efficiency''.

In summary, we have developed a comprehensive description of the
excitonic and electronic processes at type-II polymer
heterojunctions that support exciplex formation. The two blends studied here represent important examples for efficient charge generation on the one hand and high luminescence yields on the other, and in this sense represent the two extremes of type-II heterojunctions found in common semiconductor polymer blends. 
The very different behavior was shown to arise from different geminate pair separations (3.1\,nm vs. 2.2\,nm) and back-transfer activation energies (200\,meV vs. 100\,meV) which affect strongly the kinetics between the states involved. We note that both thermalization distance as well as activation energy are generally expected to be larger for larger band edge offsets between the two polymers and that this provides the link to the classification scheme described in the introduction~\cite{Halls.Cornil.ea:99}. Given that excited-state electronic dimers are commonly observed in polymeric semiconductors~\cite{Schwartz:03}, we consider exciplex formation and exciton regeneration to also be general phenomena at type II polymeric heterojunctions. As shown in this letter, the central role of these dynamics is not directly evident from steady-state PL measurements if back-transfer is efficient. \textsf{  } 

In photovoltaic operation the collapse of the geminate pair into the exciplex provides an unwanted loss channel. We suggest that optimized interfaces require not only large band-edge offsets to enable large thermalization distances ($r_{gp}$), but also inhibited exciplex stabilization. This can be achieved by increasing intermolecular distances and decreasing configurational relaxation~\cite{Weller:75}. 

\begin{acknowledgments}
This work was supported by the EPSRC. ACM is a Gates Cambridge Scholar. CS is
an EPSRC Advanced Research Fellow. We are grateful to N.C. Greenham and A.S. Dhoot for valuable discussions.
\end{acknowledgments}


\begin{thebibliography}{23}
\expandafter\ifx\csname natexlab\endcsname\relax\def\natexlab#1{#1}\fi
\expandafter\ifx\csname bibnamefont\endcsname\relax
  \def\bibnamefont#1{#1}\fi
\expandafter\ifx\csname bibfnamefont\endcsname\relax
  \def\bibfnamefont#1{#1}\fi
\expandafter\ifx\csname citenamefont\endcsname\relax
  \def\citenamefont#1{#1}\fi
\expandafter\ifx\csname url\endcsname\relax
  \def\url#1{\texttt{#1}}\fi
\expandafter\ifx\csname urlprefix\endcsname\relax\def\urlprefix{URL }\fi
\providecommand{\bibinfo}[2]{#2}
\providecommand{\eprint}[2][]{\url{#2}}

\bibitem[{\citenamefont{Halls et~al.}(1995)\citenamefont{Halls, Walsh,
  Greenham, Marseglia, Friend, Moratti, and Holmes}}]{Halls.Walsh.ea:95}
\bibinfo{author}{\bibfnamefont{J.~J.~M.} \bibnamefont{Halls}} \etal,
  \bibinfo{journal}{Nature}
  \textbf{\bibinfo{volume}{376}}, \bibinfo{pages}{498} (\bibinfo{year}{1995}).

\bibitem[{\citenamefont{Arias et~al.}(2001)\citenamefont{Arias, MacKenzie,
  Stevenson, Halls, Inbasekaran, Woo, Richards, and
  Friend}}]{Arias.MacKenzie.ea:01}
\bibinfo{author}{\bibfnamefont{A.~C.} \bibnamefont{Arias}} \etal,
  \bibinfo{journal}{Macromolecules} \textbf{\bibinfo{volume}{34}},
  \bibinfo{pages}{6005} (\bibinfo{year}{2001}).

\bibitem[{\citenamefont{Brabec et~al.}(2001)\citenamefont{Brabec, Sariciftci,
  and Hummelen}}]{Brabec.Sariciftci.ea:01}
\bibinfo{author}{\bibfnamefont{C.~J.} \bibnamefont{Brabec}},
  \bibinfo{author}{\bibfnamefont{N.~S.} \bibnamefont{Sariciftci}},
  \bibnamefont{and} \bibinfo{author}{\bibfnamefont{J.~C.}
  \bibnamefont{Hummelen}}, \bibinfo{journal}{Adv. Funct. Mater.}
  \textbf{\bibinfo{volume}{11}}, \bibinfo{pages}{15} (\bibinfo{year}{2001}).

\bibitem[{\citenamefont{Halls et~al.}(1999)\citenamefont{Halls, Cornil, dos
  Santos, Silbey, Hwang, Holmes, Br\'{e}das, and Friend}}]{Halls.Cornil.ea:99}
\bibinfo{author}{\bibfnamefont{J.~J.~M.} \bibnamefont{Halls}} \etal,
  \bibinfo{journal}{Phys. Rev. B}
  \textbf{\bibinfo{volume}{60}}, \bibinfo{pages}{5721} (\bibinfo{year}{1999}).

\bibitem[{\citenamefont{Cao et~al.}(1999)\citenamefont{Cao, Parker, Yu, Zhang,
  and Heeger}}]{Cao.Parker.ea:99}
\bibinfo{author}{\bibfnamefont{Y.}~\bibnamefont{Cao}} \etal,
  \bibinfo{journal}{Nature} \textbf{\bibinfo{volume}{397}},
  \bibinfo{pages}{414} (\bibinfo{year}{1999}).

\bibitem[{\citenamefont{Palilis et~al.}(2001)\citenamefont{Palilis, Lidzey,
  Redecker, Bradley, Inbasekaran, Woo, and Wu}}]{Palilis.Lidzey.ea:01}
\bibinfo{author}{\bibfnamefont{L.~C.} \bibnamefont{Palilis}} \etal,
  \bibinfo{journal}{Synth. Met.} \textbf{\bibinfo{volume}{121}},
  \bibinfo{pages}{1729} (\bibinfo{year}{2001}).

\bibitem[{\citenamefont{Morteani et~al.}(2003)\citenamefont{Morteani, Dooht,
  Kim, Silva, Greenham, Murphy, Moons, Cina, Burroughes, and
  Friend}}]{Morteani.Dooht.ea:03}
\bibinfo{author}{\bibfnamefont{A.~C.} \bibnamefont{Morteani}} \etal,
  \bibinfo{journal}{Adv. Mat.} \textbf{\bibinfo{volume}{15}}, \bibinfo{pages}{1708} (\bibinfo{year}{2003}).

\bibitem[{\citenamefont{Alvarado et~al.}(1998)\citenamefont{Alvarado, Seidler,
  Lidzey, and Bradley}}]{Alvarado.Seidler.ea:98}
\bibinfo{author}{\bibfnamefont{S.}~\bibnamefont{Alvarado}} \etal,
  \bibinfo{journal}{Phys. Rev. Lett.} \textbf{\bibinfo{volume}{81}},
  \bibinfo{pages}{1082} (\bibinfo{year}{1998}).

\bibitem[{\citenamefont{Weller}(1975)}]{Weller:75}
\bibinfo{author}{\bibfnamefont{A.}~\bibnamefont{Weller}}, in
  \emph{\bibinfo{booktitle}{The Exciplex}}, edited by
  \bibinfo{editor}{\bibfnamefont{M.}~\bibnamefont{Gordon}} \bibnamefont{and}
  \bibinfo{editor}{\bibfnamefont{W.}~\bibnamefont{Ware}}
  (\bibinfo{publisher}{Academic Press Inc.}, \bibinfo{address}{New York},
  \bibinfo{year}{1975}), pp. \bibinfo{pages}{23--38}.

\bibitem[{\citenamefont{Snaith et~al.}(2002)\citenamefont{Snaith, Arias,
  Morteani, Silva, and Friend}}]{Snaith.Arias.ea:02}
\bibinfo{author}{\bibfnamefont{H.~J.} \bibnamefont{Snaith}} \etal,
  \bibinfo{journal}{Nano Lett.} \textbf{\bibinfo{volume}{2}},
  \bibinfo{pages}{1353} (\bibinfo{year}{2002}).

\bibitem[{\citenamefont{Hayes et~al.}(1995)\citenamefont{Hayes, Samuel, and
  Phillips}}]{Hayes.Samuel.ea:95}
\bibinfo{author}{\bibfnamefont{G.~R.} \bibnamefont{Hayes}} \etal, 
  \bibinfo{journal}{Phys. Rev. B}
  \textbf{\bibinfo{volume}{52}}, \bibinfo{pages}{11569} (\bibinfo{year}{1995}).

\bibitem{LanzaniPaper}
We note that field-assisted exciton dissociation has been seen in a related polyfluorene at higher fields
(see \bibinfo{author}{\bibfnamefont{T.} \bibnamefont{Virgili}} \etal, 
 \bibinfo{journal}{Phys. Rev. Lett.}
 \textbf{\bibinfo{volume}{90}}, \bibinfo{pages}{247402} (\bibinfo{year}{2003})).

\bibitem[{\citenamefont{Yokoyama et~al.}(1981)\citenamefont{Yokoyama, Endo,
  Matsubara, and Mikawa}}]{Yokoyama.Endo.ea:81}
\bibinfo{author}{\bibfnamefont{M.}~\bibnamefont{Yokoyama}} \etal,
  \bibinfo{journal}{J. Chem. Phys.} \textbf{\bibinfo{volume}{75}},
  \bibinfo{pages}{3006} (\bibinfo{year}{1981}).

\bibitem[{\citenamefont{Wu and Conwell}(1998)}]{Wu.Conwell:98}
\bibinfo{author}{\bibfnamefont{M.~W.} \bibnamefont{Wu}} \bibnamefont{and}
  \bibinfo{author}{\bibfnamefont{E.~M.} \bibnamefont{Conwell}},
  \bibinfo{journal}{Chem. Phys.} \textbf{\bibinfo{volume}{227}},
  \bibinfo{pages}{11} (\bibinfo{year}{1998}).

\bibitem[{\citenamefont{Ohta}(2002)}]{Ohta:02}
\bibinfo{author}{\bibfnamefont{N.}~\bibnamefont{Ohta}}, \bibinfo{journal}{Bull.
  Chem. Soc. Jpn.} \textbf{\bibinfo{volume}{75}}, \bibinfo{pages}{1637}
  (\bibinfo{year}{2002}).

\bibitem[{\citenamefont{Arkhipov et~al.}(2003)\citenamefont{Arkhipov, Heremans,
  and B\"{a}ssler}}]{Arkhipov.Heremans.ea:03}
\bibinfo{author}{\bibfnamefont{V.~I.} \bibnamefont{Arkhipov}},
  \bibinfo{author}{\bibfnamefont{P.}~\bibnamefont{Heremans}}, \bibnamefont{and}
  \bibinfo{author}{\bibfnamefont{H.}~\bibnamefont{B\"{a}ssler}},
  \bibinfo{journal}{Appl. Phys. Lett.} \textbf{\bibinfo{volume}{82}},
  \bibinfo{pages}{4605} (\bibinfo{year}{2003}).

\bibitem[{\citenamefont{Nikitenko et~al.}(2001)\citenamefont{Nikitenko, Hertel,
  and B\"{a}ssler}}]{Nikitenko.Hertel.ea:01}
\bibinfo{author}{\bibfnamefont{V.~R.} \bibnamefont{Nikitenko}},
  \bibinfo{author}{\bibfnamefont{D.}~\bibnamefont{Hertel}}, \bibnamefont{and}
  \bibinfo{author}{\bibfnamefont{H.}~\bibnamefont{B\"{a}ssler}},
  \bibinfo{journal}{Chem. Phys. Lett.} \textbf{\bibinfo{volume}{348}},
  \bibinfo{pages}{89} (\bibinfo{year}{2001}).

\bibitem[{\citenamefont{Pope and Swenberg}(1999)}]{Pope.Swenberg:99}
\bibinfo{author}{\bibfnamefont{M.}~\bibnamefont{Pope}} \bibnamefont{and}
  \bibinfo{author}{\bibfnamefont{C.~E.} \bibnamefont{Swenberg}},
  \emph{\bibinfo{title}{Electronic Processes in Organic Crystals and Polymers}}
  (\bibinfo{publisher}{Oxford University Press}, \bibinfo{year}{1999}),
  \bibinfo{edition}{2nd} ed.

\bibitem{OnsagerEquations}
$f_{\epsilon}\left(r_{gp}, T,
F\right)=\frac{1}{2}\int^{\pi}_{0}d\theta\, sin\,\theta\, e^{-(A+B)} \sum_{m,n=0}^{\infty}\,\frac{A^{m}}{m!}\frac{B^{m+n}}{(m+n)!}$; $A=2q/r_{gp}$, $B=\beta r_{gp} (1+cos\,\theta)$, $q=e^{2}/8\pi\epsilon\epsilon_{0}kT$, $\beta=eF/2kT$ (see ref.~\cite{Pope.Swenberg:99}, p.~484). Averaging over $\theta$ represents an isotropic blend morphology. 

\bibitem{InternalField}
Internal field calculated as $F=(\text{applied Voltage}+x)/170\,\text{nm}$; $x$ is the Voltage corresponding to minimum $-\Delta PL/PL$, found to be 0.7\,V (PFB:F8BT) and 0.3\,V (TFB:F8BT) forward bias.

\bibitem{TemperatureArgument}
At temperatures below 230\,K the $\Delta$PL arising from the excitons' Stark-shift obscures the weak quenching signal.





\bibitem[{\citenamefont{Schwartz}(2003)}]{Schwartz:03}
\bibinfo{author}{\bibfnamefont{B.~J.} \bibnamefont{Schwartz}},
  \bibinfo{journal}{Annu. Rev. Phys. Chem.} \textbf{\bibinfo{volume}{54}},
  \bibinfo{pages}{141} (\bibinfo{year}{2003}) and references therein.

\end{thebibliography}
\end{document}